\documentclass[final]{aipproc}
\layoutstyle{6x9}
\usepackage{xcolor}
\pdfoutput=1

\begin{document}

\vspace*{-15mm}
\begin{flushright}
MPP-2009-150\\
\end{flushright}
\vspace*{0.2cm}

\title{GUT predictions for quark and lepton\\ mass ratios}

\classification{12.10.Kt, 12.60.Jv}
\keywords      {SUSY GUTs, Yukawa coupling unification}

\author{S. Antusch}{
  address={Max-Planck-Institut f\"ur Physik (Werner-Heisenberg-Institut),\\
  F\"ohringer Ring 6, D-80805 M\"unchen, Germany}
}

\author{M. Spinrath}{
}

\begin{abstract}
Group theoretical factors from GUT symmetry breaking can lead to predictions for the ratios of quark and lepton masses at the unification scale. Due to supersymmetric (SUSY) threshold corrections the viability of such predictions can depend strongly on the SUSY parameters. We derive possible new predictions for the GUT scale ratios $m_\mu/m_s$, $y_\tau/y_b$ and $y_t/y_b$ and compare them with the experimentally allowed ranges for three common SUSY breaking scenarios.
\end{abstract}

\maketitle

\section{Motivation}

The flavour sector contains thirteen of the nineteen free parameters of the Standard Model (SM). Adding neutrino masses to the SM introduces on top of that at least seven additional parameters. Such a large number of parameters seems not appropriate for a fundamental theory. These parameters furthermore exhibit an unexpected pattern. The masses are strongly hierarchical and the mixing angles in the quark sector are small, whereas two of the three mixing angles in the lepton sector are large.

There are various attempts to reduce the number of parameters in the flavour sector. One very popular attempt is to introduce additional symmetries like flavour symmetries. We will focus here on models of (SUSY) Grand Unified Theories (GUTs), where some of the Yukawa couplings can be related to each other due to the underlying group structure.

\section{GUT predictions for mass ratios}

In GUTs joined representations contain different kinds of fermions. GUT invariant operators, which generate the low energy Yukawa couplings after symmetry breaking, therefore typically relate some of the Yukawa couplings to each other. An operator like $\mathbf{5}_F \bar{\mathbf{10}}_F \langle \bar{\mathbf{5}}_H \rangle$ in SU(5), where $\mathbf{5}_F$ stands for the five dimensional matter, $\bar{\mathbf{10}}_F$ for the ten dimensional matter and $\langle \bar{\mathbf{5}}_H \rangle$ for the vacuum expectation value (vev) of the five dimensional Higgs representation, generates after SU(5) breaking identical GUT scale Yukawa couplings for down type quarks and charged leptons. This kind of operator generates $b$-$\tau$-unification of Yukawa couplings. If the vev of the five dimensional representation is replaced by the vev of a 45 dimensional representation, the ratio is changed to $y_\mu/y_s = -3$. This relation is the famous Georgi Jarlskog (GJ) relation \cite{GJ}.

In \cite{Antusch:2009gu} Yukawa coupling ratios predicted from dimension four and five operators in SU(5) and Pati-Salam (PS) gauge theory are studied systematically. For this the following three assumptions have been made:
\begin{enumerate}
 \item The Yukawa matrices in the flavour basis are hierarchical and dominated by the diagonal elements. This is reasonable for the second and third generation and in this way the ratio of the masses directly corresponds to the ratio of the (2,2) resp. (3,3) entries of the Yukawa matrices, which can be predicted from the underlying group structure.
 \item The (2,2) and (3,3) elements are dominated by one single operator. Otherwise different ratios mix up and the predicitons are obscured.
 \item All the fields are included within common SO(10) models, so that an embedding in SO(10) is possible.
\end{enumerate}
In the conclusions we will comment on which relations are viable within our setup.

\section{SUSY threshold corrections}

To relate GUT scale Yukawa coupling ratios to low energy fermion masses within a SUSY theory it is very important to include SUSY threshold corrections in the renormalization group running, especially for large $\tan \beta$. The matching formula for the down type quarks and charged leptons is
\begin{eqnarray*}
y_i^{MSSM} &=& \frac{y_i^{SM}}{\cos \beta (1 + \epsilon_i \tan \beta)} \;,
\end{eqnarray*}
where $i$ stands for $d$, $s$, $b$, $e$, $\mu$ and $\tau$ and $\epsilon_i$ is a one loop factor. Explicit expressions for $\epsilon_i$ can be found in \cite{Antusch:2008tf}, where the importance of SUSY threshold corrections for the GUT scale Yukawa coupling ratios in the electroweak unbroken phase are analyzed.

\begin{table}
\caption{Ranges for the GUT scale ratios $m_\mu/m_s$, $m_e/m_d$, $y_\tau/y_b$ and $y_t/y_b$ corresponding to example ranges of SUSY parameters $g_+$, $g_-$ and $a$ defined in \cite{Antusch:2008tf} including $A_t = 0$ for $\tan \beta = 30$. Case 0 refers to the case without SUSY threshold corrections. The present experimental errors for the quark masses have been included.\label{tab:Ratfinal} }
\centering
\begin{tabular}{ccccc}\hline
 \tablehead{1}{c}{b}{Ratio}	& \tablehead{1}{c}{b}{Case 0} & \tablehead{1}{c}{b}{Case $g_+$} & \tablehead{1}{c}{b}{Case $g_-$} & \tablehead{1}{c}{b}{Case $a$} \\ \hline
 $m_e/m_d$	& [0.28, 0.67] & [0.30, 0.86] & [0.21, 0.61] & [0.20, 0.62] \\
 $m_\mu/m_s$	& [3.39, 6.07] & [3.73, 7.79] & [2.54, 5.49] & [2.40, 5.63] \\
 $y_\tau/y_b$	& [1.27, 1.38] & [1.20, 2.02] & [0.71, 1.43] & [0.60, 1.39] \\
 $y_t/y_b$	& [2.56, 3.02] & [2.36, 4.19] & [1.50, 3.28] & [1.14, 2.87] \\ \hline
\end{tabular}
\end{table}

In table \ref{tab:Ratfinal} some exemplary results are shown. For case $g_+$, which is inspired by unification of gaugino masses with positive $\mu$ the ratios are shifted to larger values compared to case 0, where no thresholds are included. For case $g_-$ and $a$, which are inspired by unification of gaugino masses with negative $\mu$ and anomaly mediated SUSY breaking scenarios, the ratios are shifted to smaller values. So the SUSY spectrum resp. the SUSY breaking parameters and especially their signs have a strong impact on the GUT scale mass ratios via SUSY threshold corrections.

\section{Experimental constraints}

For the calculation of the complete SUSY spectrum a modified version of SoftSUSY 2.0.18 \cite{SoftSUSY} was used, where the full SUSY threshold corrections and left right mixing of sfermions in the electroweak unbroken phase for the first two generations were included. Via the SLHA \cite{SLHA} interface the spectra are imported to micrOMEGAs 2.2 CPC \cite{micromegas} for the calculation of further observables, which served as constraints.

For the results shown in figure \ref{fig:finalresults} we have used the following constraints:
\begin{itemize}
\item Direct detection: Masses of $\chi^\pm$, $\tilde{l}$ and $\tilde{\nu}$ heavier than LEP bounds  
\item Electroweak precision observables:
\begin{itemize}
 \item $M_W = 80.429 \pm 0.039$~GeV
 \item $\sin^2 \theta_{eff} = 0.23153 \pm 0.00016$
\end{itemize}
\item B-physics:
\begin{itemize}
 \item $BR\left(b\rightarrow s \gamma\right) = \left(3.55^{+0.36}_{-0.37}\right) \times 10^{-4}$
 \item $BR\left(B_s \rightarrow \mu^+ \mu^-\right) \leq 5.8 \times 10^{-8}$
\end{itemize}
\item Anomalous magnetic moment of the muon: $ 0 \leq a_\mu \leq 35.9 \times 10^{-10}$
\end{itemize}
For a detailed discussion of these constraints and the discussion of an additional dark matter constraint see \cite{Antusch:2009gu}.

\section{Summary and Conclusions}

\begin{table}
\caption{Summary of the results. A checkmark (red cross) denotes (in)compatibility within the experimental constraints. \label{tab:summary}}
\centering
\begin{tabular}{ccccc}\hline
\tablehead{1}{c}{b}{Prediction} & \tablehead{1}{c}{b}{Gauge group} & \tablehead{1}{c}{b}{mAMSB} & \tablehead{1}{c}{b}{mGMSB} & \tablehead{1}{c}{b}{CMSSM} \\ \hline
$m_\mu/m_s = 3$      & SU(5), PS & $\surd$ & {\color{red} $\times$} & {\color{red} $\times$}  \\
$y_t = y_b = y_\tau$ & SU(5), PS & {\color{red} $\times$} & {\color{red} $\times$} & {\color{red} $\times$} \\
$y_\tau/y_b = 1$\tablenote{see comment in footnote 1}     & SU(5), PS & $\surd$ & {\color{red} $\times$} & {\color{red} $\times$}  \\
$m_\mu/m_s = 6$      & SU(5)     & {\color{red} $\times$} & $\surd$ & $\surd$  \\
$m_\mu/m_s = 9/2$    & SU(5)     & $\surd$ & $\surd$ & $\surd$  \\
$y_\tau/y_b = 3/2$   & SU(5)     & {\color{red} $\times$} & $\surd$ & $\surd$  \\
$2 y_t = 2 y_b = y_\tau$\tablenote{taken from \cite{Allanach:1996hz}}  & PS & {\color{red} $\times$} & $\surd$ & $\surd$  \\
$y_t = 2 y_b = 2 y _\tau$$^\dagger$ & PS & $\surd$ & {\color{red} $\times$} & {\color{red} $\times$} \\
\hline
\end{tabular}
\end{table}

\begin{figure}
 \centering
 \includegraphics[scale=0.35]{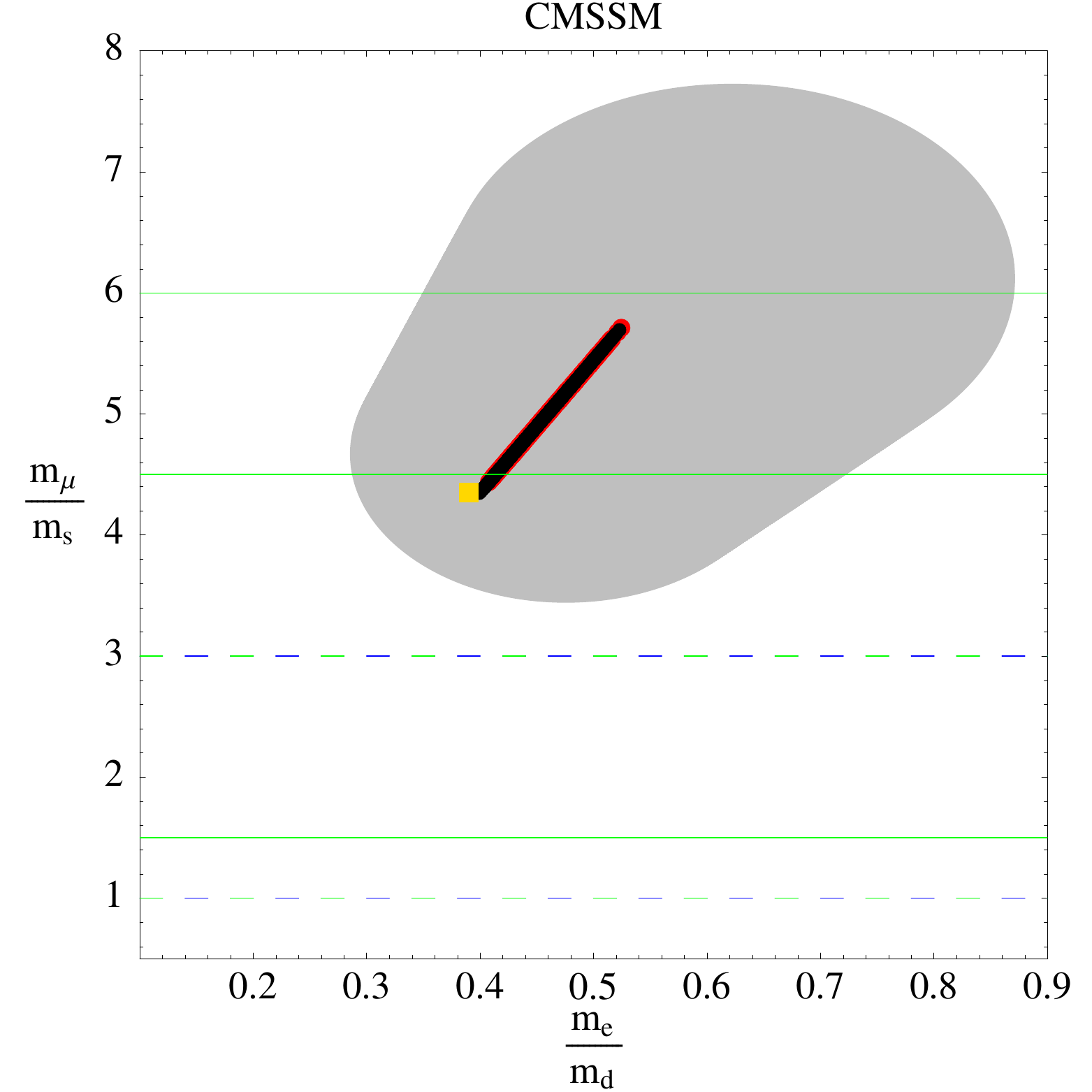} \hspace{1cm}
 \includegraphics[scale=0.35]{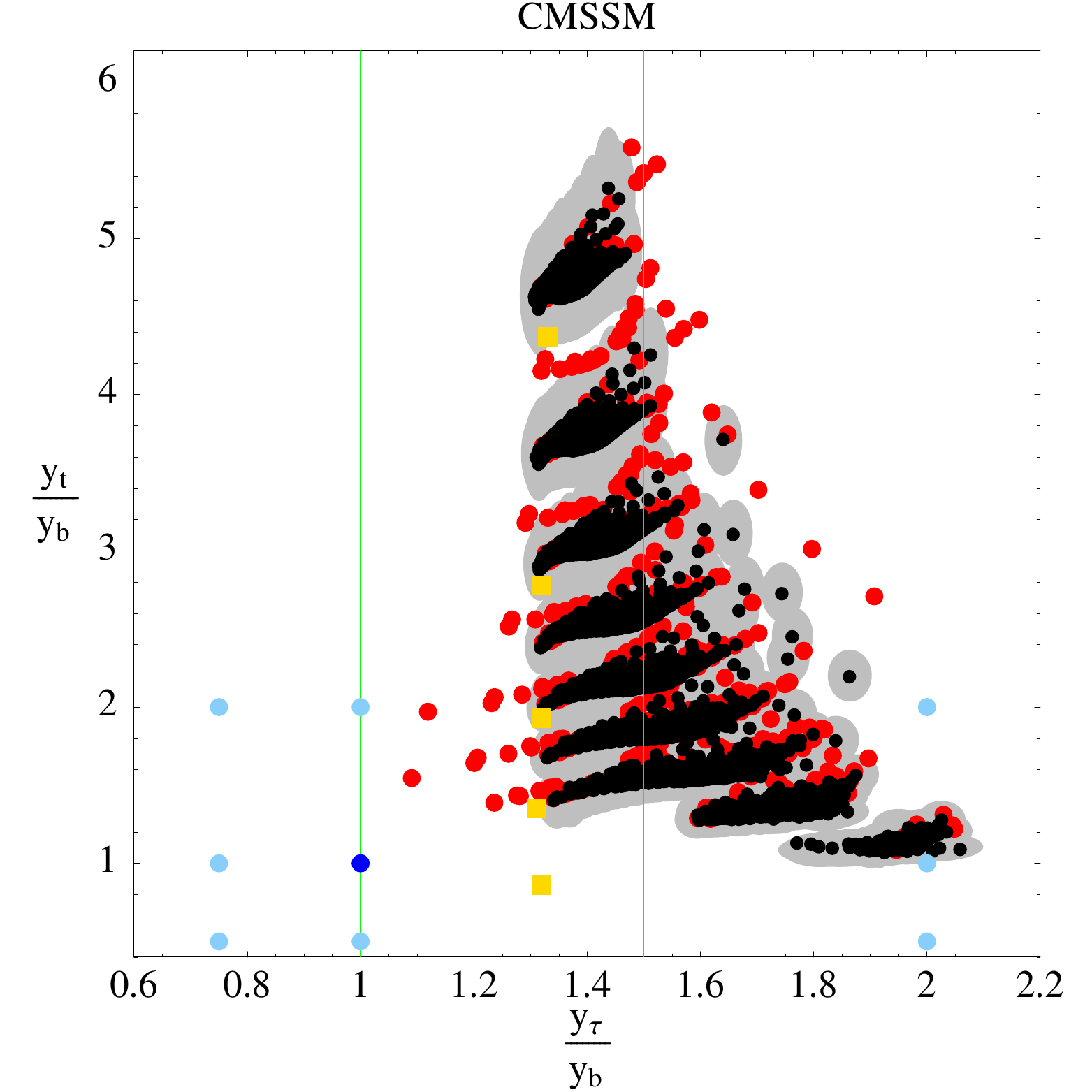} 
 \caption{Final results for mAMSB, mGMSB and CMSSM. 
          The (red) black points are the (excluded) allowed points after
          applying the constraints. The grey regions indicate the uncertainties 
	  from experimental quark mass errors. 
	  The green lines are predictions from SU(5), the dashed lines from SU(5) and PS 
	  and the (light) blue points from PS (dimension-six operators). 
	  The yellow squads are the GUT scale Yukawa ratios without including
          SUSY threshold corrections for
          $\tan \beta = 20$, $30$, $40$, $50$ and $60$ from top to bottom.
   \label{fig:finalresults}}
\end{figure}

The first thing to notice about our results summarized in table \ref{tab:summary} is, that the common GUT assumptions of the GJ relation and complete third family Yukawa coupling unification are challenged in our setup. Whereas the GJ relation is possible only for the mAMSB scenario for $\tan \beta \sim 30$, complete third family Yukawa coupling unification is incompatible with our results in all three scenarios (see also figure \ref{fig:finalresults}). In the mAMSB case, where the corrections point in the right direction to achieve this relation without fine tuning, unification cannot be achieved because of non successful electroweak symmetry breaking and tachyons in the spectrum.

Nevertheless we find valid alternatives to the standard assumptions. For the second generation we find the SU(5) relations $y_\mu/y_s = 9/2$ for $\tan \beta \sim 20$ and $y_\mu/y_s = 6$ for $\tan \beta \sim 50$ (not for mAMSB). The first of these is even valid for all three SUSY breaking scenarios (and smaller $\tan \beta$), which is simply due to the fact, that it emerges when the threshold corrections are small and therefore the dependence on the spectrum is weak.

For the third generation the SU(5) and PS relation $y_\tau/y_b = 1$ can be realised in mAMSB for $\tan \beta \sim 30$.\footnote{In our study we have focused on $\tan \beta$ between 20 and 60. We note that there can also be $b$-$\tau$-unification of Yukawa couplings for $\tan \beta \sim 1$ (see e.\ g.\ \cite{Ross:2007az}).} The SU(5) relation $y_\tau/y_b = 3/2$ appears in mGMSB for $\tan \beta \sim 30$-$45$ and in CMSSM for $\tan \beta \sim 20$-$50$. There are also alternatives in PS models, namely the relations $2 y_t = 2 y_b = y_\tau$ for $\tan \beta \sim 60$ in mGMSB and CMSSM and $y_t = 2 y_b = 2 y_\tau$ for $\tan \beta \sim 30$ in mAMSB. These relations are however generated from dimension six operators and have been obtained in \cite{Allanach:1996hz}, where a slightly different setup has been used.

In summary we found that in our setup new GUT predictions for quark and lepton mass ratios such as $m_\mu/m_s = 9/2$ or $6$ and $y_\tau/y_b = 3/2$ or $2$ are often favoured compared to the ubiquitous relations $m_\mu/m_s = 3$ and $y_\tau/y_b = 1$. These alternative relations point to characteristic SUSY spectra and breaking mechanisms which can be tested at the CERN LHC and future colliders.

\begin{theacknowledgments}
This work is partially supported by 
the DFG cluster of excellence ``Origin and Structure of the Universe''.
\end{theacknowledgments}

\bibliographystyle{aipproc}   % if natbib is available

\newpage

\begin{figure}
 \centering
 \includegraphics[scale=0.35]{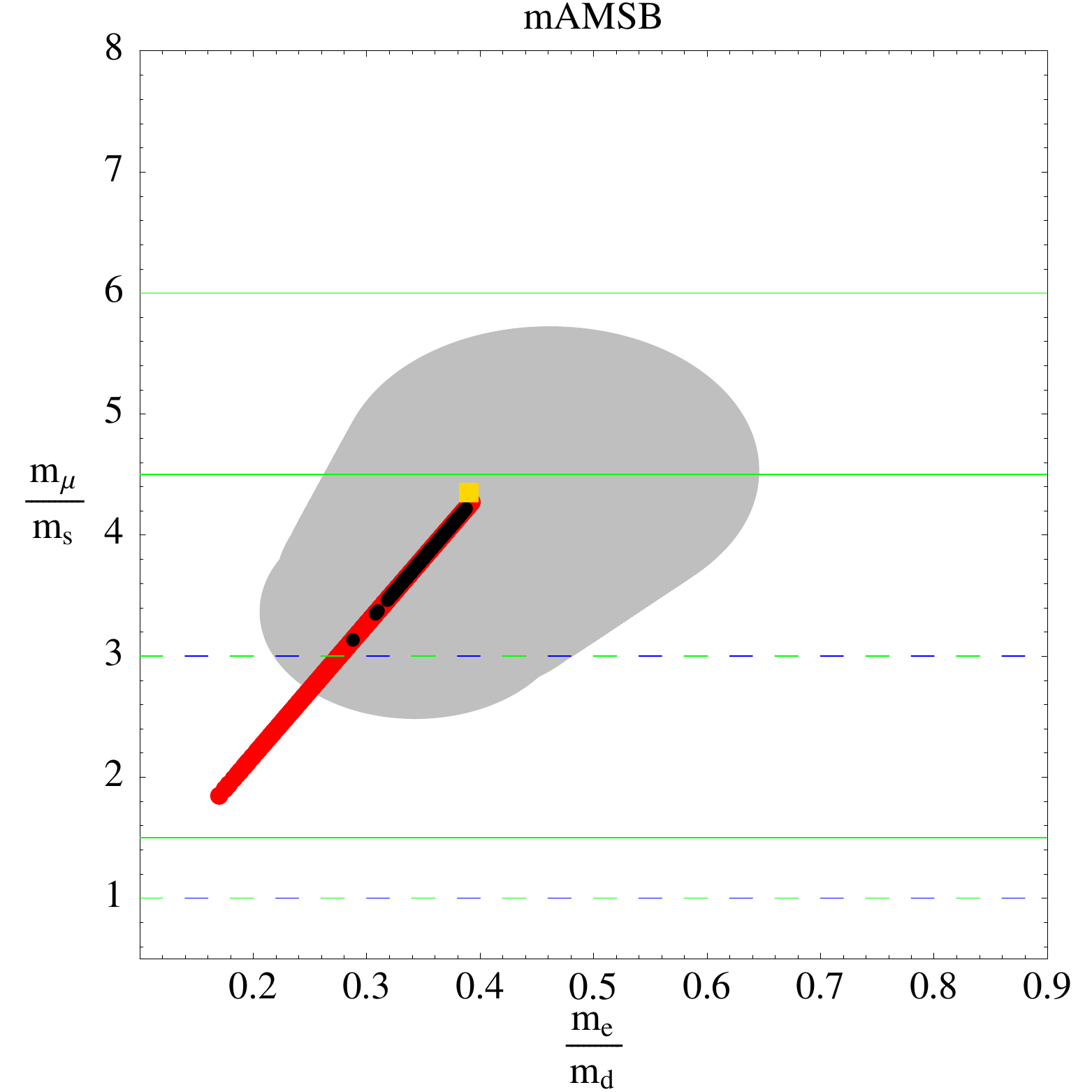} \hspace{1cm}
 \includegraphics[scale=0.35]{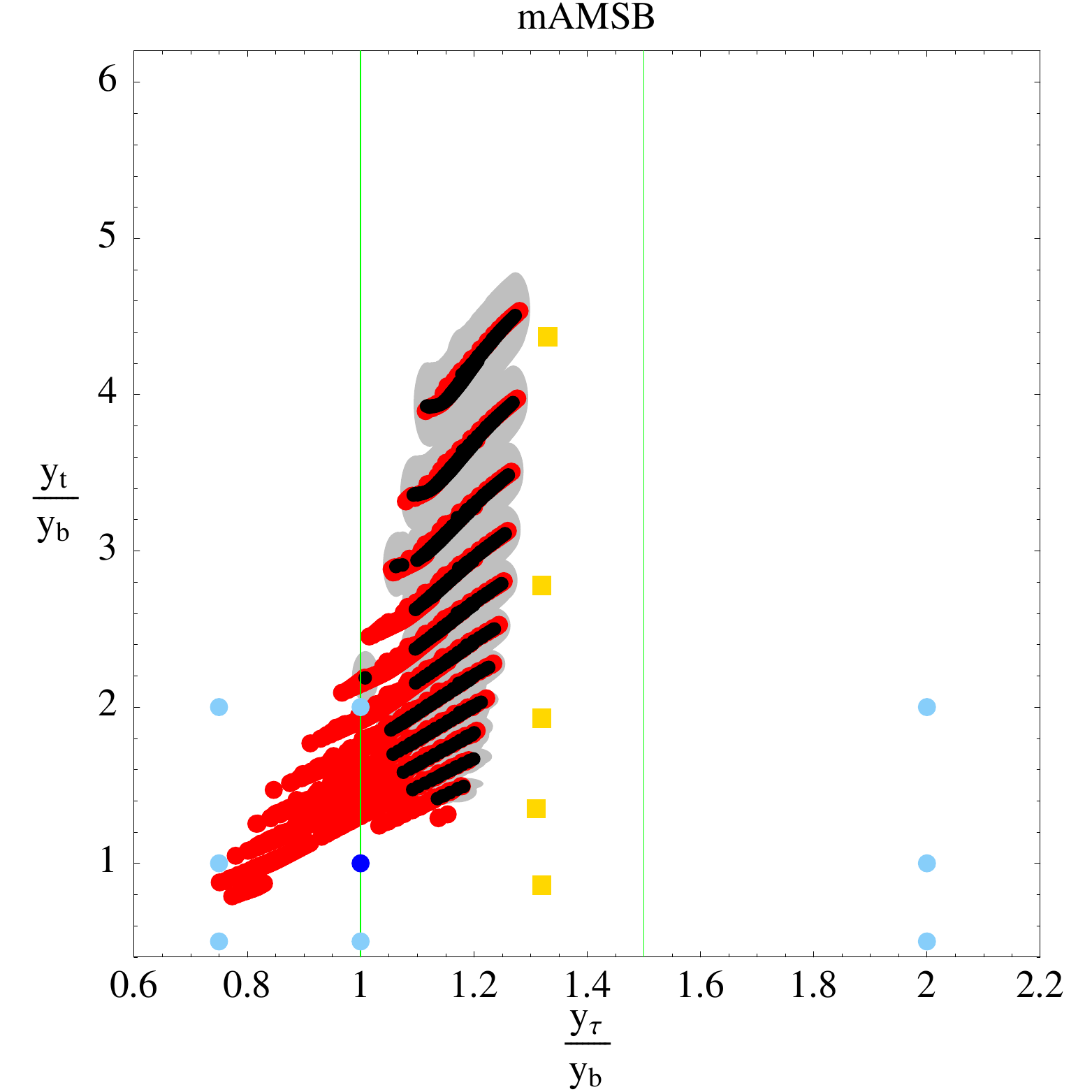} 
\end{figure}
\begin{figure}
 \centering
 \includegraphics[scale=0.35]{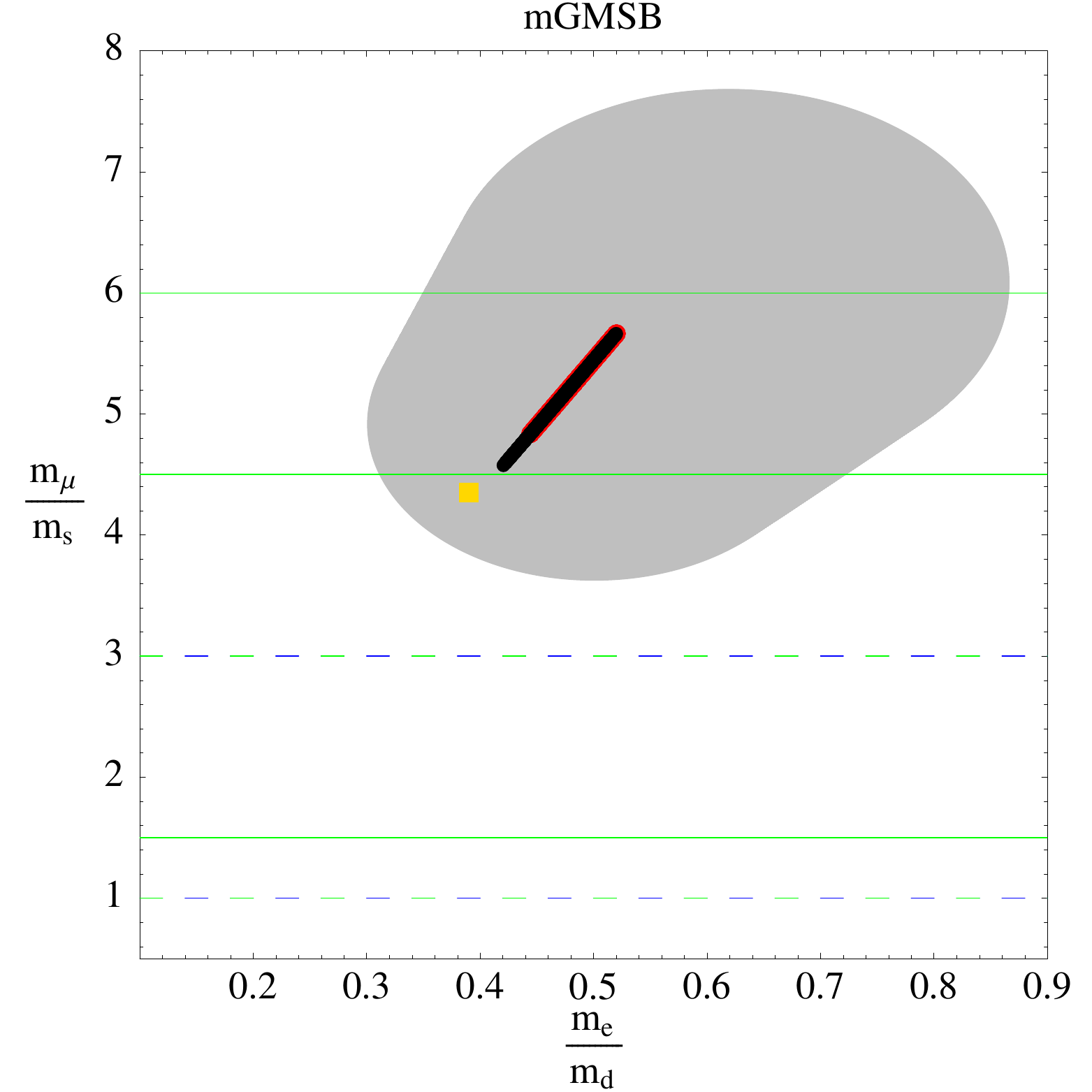} \hspace{1cm}
 \includegraphics[scale=0.35]{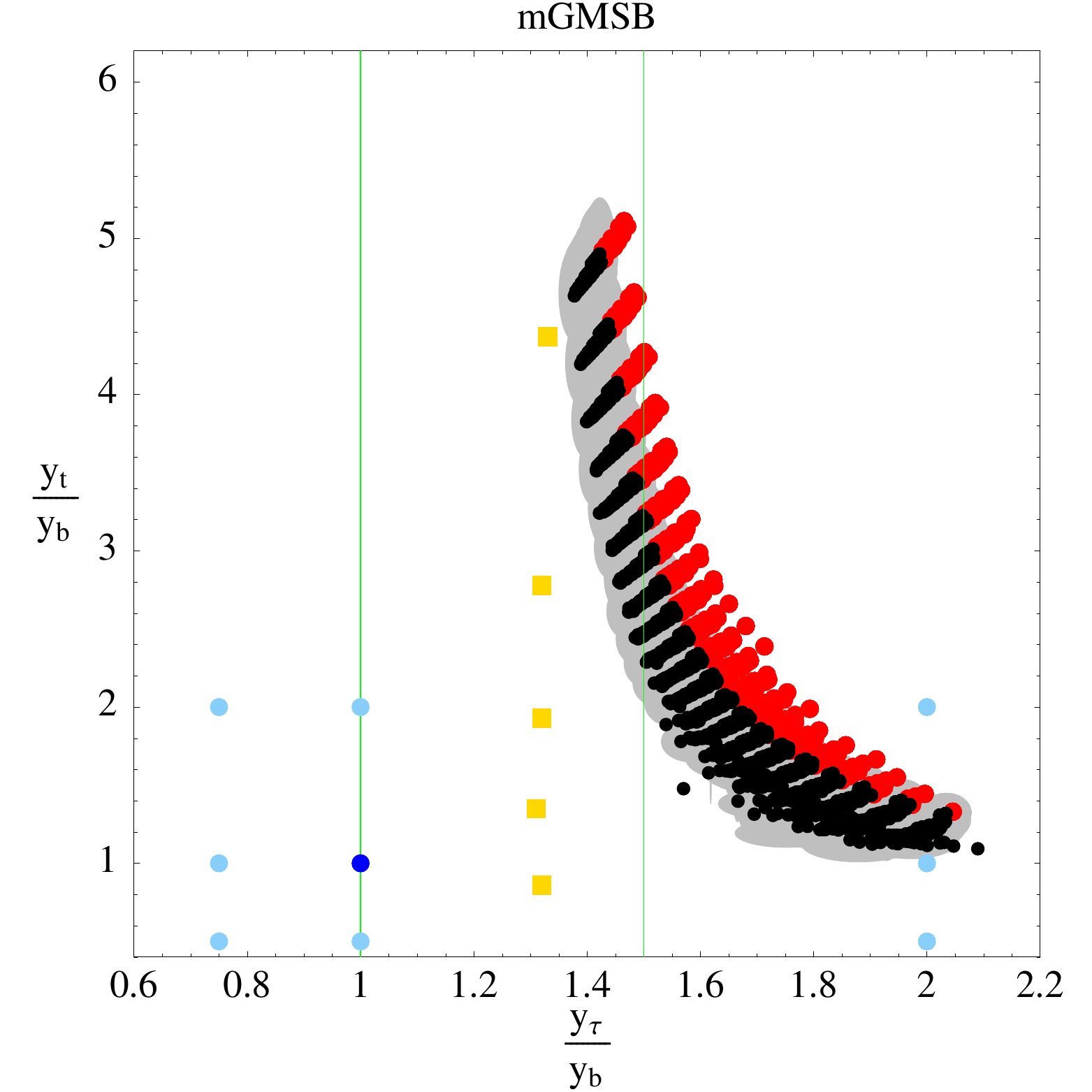} 
\end{figure}

\end{document}